\documentclass{jfm}
\usepackage{graphicx}
\usepackage{epstopdf, epsfig}
\usepackage{xcolor}

\usepackage{enumerate}
\usepackage{enumitem}
\usepackage{amsmath}
\usepackage{esint}
\usepackage{dutchcal}
\usepackage{bbm}
\shorttitle{Energy cascade in the Garrett-Munk spectrum}
\shortauthor{Yue Wu and Yulin Pan}

\title{Energy cascade in the Garrett-Munk spectrum of internal gravity waves}

\author{Yue Wu\aff{1}
 \corresp{\email{ywu.ocean@gmail.com}}
 \and Yulin Pan\aff{1}}

\affiliation{\aff{1}Naval Architecture and Marine Engineering, University of Michigan, Ann Arbor, MI, USA}

\begin{document}
\maketitle

\begin{abstract}
We study the spectral energy transfer due to wave-triad interactions in the Garrett-Munk spectrum of internal gravity waves (IGWs) based on a numerical evaluation of the collision integral in the wave kinetic equation. Our numerical evaluation builds on the reduction of the collision integral on the resonant manifold for a horizontally isotropic spectrum. We directly evaluate the downscale energy flux available for ocean mixing, whose value is in close agreement with the finescale parameterization. We further decompose the energy transfer into contributions from different mechanisms, including local interactions and three types of nonlocal interactions, namely parametric subharmonic instability (PSI), elastic scattering (ES) and induced diffusion (ID). Through analysis on the role of each mechanism, we resolve two long-standing paradoxes regarding the mechanism for forward cascade in frequency and zero ID flux for GM76 spectrum. In addition, our analysis estimates the contribution of each mechanism to the energy transfer in each spectral direction and reveals new understanding of the importance of local interactions and ES in the energy transfer.
\end{abstract}


\section{Introduction}

Internal gravity waves (IGWs) are ubiquitous features of the ocean but are filtered out by the quasi-geostrophic description of the system. Although they generally account for only a small fraction of the kinetic energy of the overall ocean, their existence has profound physical significance: they play an important role in transferring momentum, heat and tracers across the ocean, and their breaking drives most of the turbulence that leads to the inhomogeneity of ocean mixing which in turn affects the large-scale circulation.

In the IGW continuum, energy is supplied at large scales by atmospheric and tidal forcings and is dissipated at small scales. Understanding the energy transfer across scales driven by nonlinear processes is one topic of central importance in physical oceanography. Such understanding will not only shed light on the physical interpretation of IGW spectral forms, generally considered as the Garrett-Munk spectrum (GM) and its variations \citep{Garrett1972, Garrett1975, Cairns1976}, but also provide an estimate of the energy flux toward small dissipation scales (downscale energy flux) that is available for ocean mixing. The latter aspect, as one focus of our current work, is particularly important given the fact that information at small dissipation scales is difficult to obtain from both measurements and modeling.

While the downscale energy cascade of the IGW continuum can be excited by many mechanisms such as wave-eddy interactions (e.g., \citealt{Watson1985}) and bottom scattering (e.g., \citealt{Kunze2004}), in many cases, nonlinear wave interactions are considered as a major contributor to the cascade in abyssal oceans (e.g., \citealt{Muller1986,Polzin2011}). 
In quantification of spectral energy transfer due to nonlinear wave interactions, one critical tool is the wave kinetic equation (WKE) derived in the framework of wave turbulence theory \citep{Zakharov1992, Nazarenko2011}. For systems with three-wave resonant interactions, the general form of WKE reads
\begin{eqnarray} \label{eq_KE}
\frac{\partial n}{\partial t} &=\iint 4\pi |V(\mathbf{p},\mathbf{p}_1,\mathbf{p}_2)|^2f_{p12}\delta(\omega-\omega_1-\omega_2)\delta(\mathbf{p}-\mathbf{p}_1-\mathbf{p}_2) \, d\mathbf{p}_1 \, d\mathbf{p}_2\\ \nonumber
&-\iint 8\pi |V(\mathbf{p}_1,\mathbf{p},\mathbf{p}_2)|^2f_{1p2}\delta(\omega-\omega_1+\omega_2)\delta(\mathbf{p}-\mathbf{p}_1+\mathbf{p}_2) \, d\mathbf{p}_1 \, d\mathbf{p}_2, 
\end{eqnarray}
where $n(\mathbf{p},t)$ is spectral action density with $\mathbf{p}$ being the vector of wavenumber and $t$ the time, $V$ is the interaction coefficient, $\omega$ is wave frequency, $f_{p12}= n_1n_2 -n_p(n_1+n_2)$ and $f_{1p2}= n_pn_2 -n_1(n_p+n_2)$. The right-hand-side (RHS) of \eqref{eq_KE} is also referred to as the collision integral, which describes wave action evolution due to triad interactions. Other mechanisms such as generation/dissipation of IGWs and coupling with eddies/currents are not included. The existence of eddies and currents may be potentially important in nonlinear energy transfer (e.g., \citealt{Kafiabad2019,Savva2021,Dong2020,Dong2023}) but will not be the focus of the current work. For IGWs, $\mathbf{p}=(k_x,k_y,m)$, $\omega^2=(N^2 k^2+f^2 m^2)/(k^2+m^2)$ is the dispersion relation with $k=(k^2_x+k^2_y)^{1/2}$ being the magnitude of horizontal wavenumbers, $N$ the buoyancy frequency, and $f$ the Coriolis frequency.
The interaction coefficient $V$ has been derived using various approaches in the literature (e.g., \citealt{Olbers1974,Muller1975a,Olbers1976,McComas1977a,Lvov2001,Lvov2004b,Lvov2010}), which yield formulations shown to be equivalent on the resonant manifold (\citealt{Lvov2012}). 

The collision integral has nonzero values on the resonant manifold
\begin{equation}
\mathbf{p}=\mathbf{p}_1 \pm \mathbf{p}_2 \text{~~~~and~~~~} \omega=\omega_1\pm\omega_2.
\end{equation}
Therefore, \eqref{eq_KE} provides the energy transfer rate through collections of triad interactions in the spectral space. Such WKE characterizes the spectral evolution in the kinetic (or nonlinear) time scale $\tau_\mathbf{p}^{NL}$ and is valid only for weakly nonlinear waves whose linear time scale $\tau_\mathbf{p}^{L}$ ($=2\pi/\omega_\mathbf{p}$, i.e., wave period) is (much) smaller than $\tau_\mathbf{p}^{NL}$, that is, the normalized Boltzmann rate \citep{Nazarenko2011, Lvov2012}
\begin{equation}
|\varepsilon_\mathbf{p}|=\frac{\tau_\mathbf{p}^L}{\tau_\mathbf{p}^{NL}}=\left| \frac{2\pi (\partial n_\mathbf{p}/\partial t)}{\omega_\mathbf{p} n_\mathbf{p}} \right| \ll 1. \label{eq_Bo}
\end{equation}

\begin{figure}
 \centerline{\includegraphics[width=.8\textwidth]{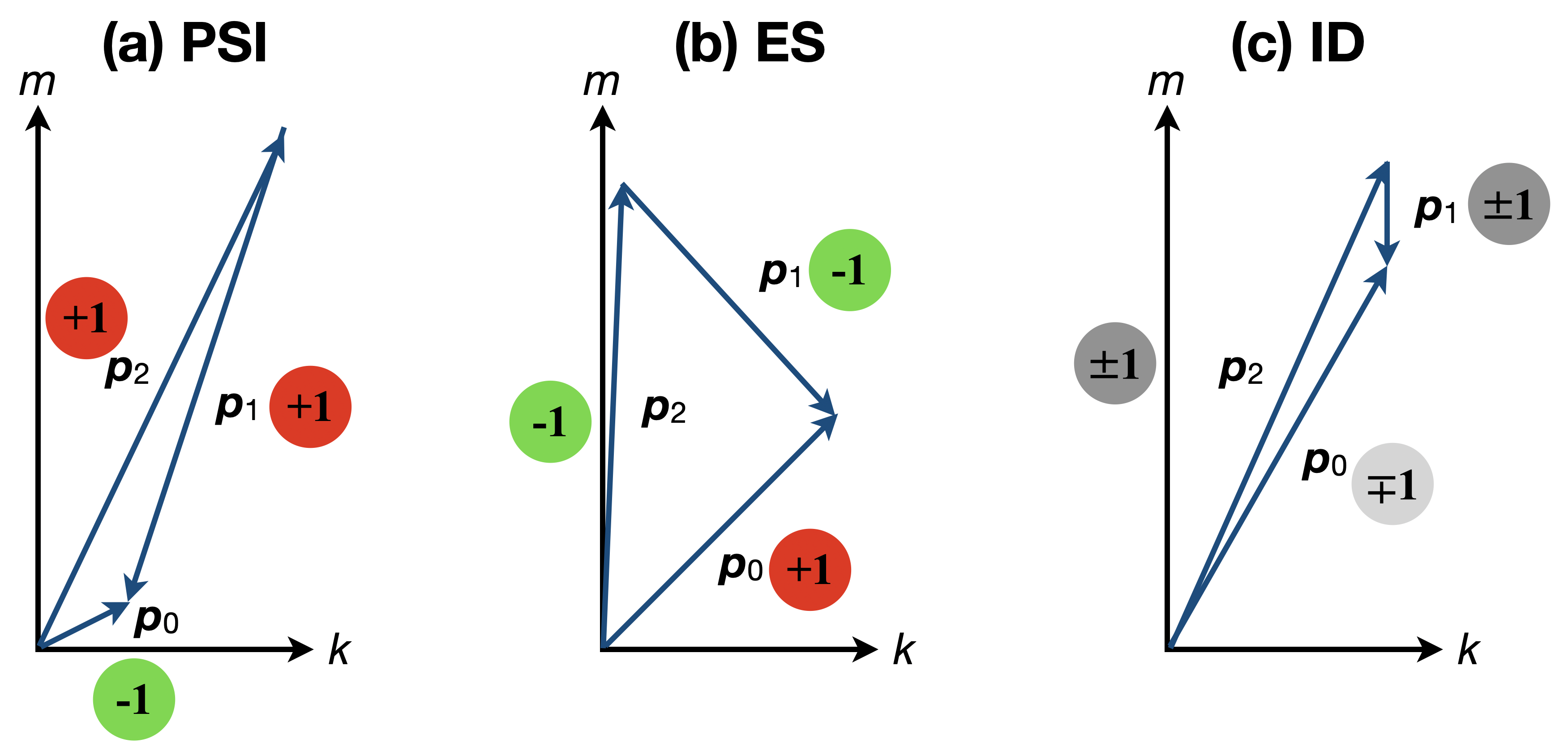}}
 \caption{The wavenumber vectors and action balance for the three types of scale-separated interactions with 
 $\mathbf{p}_0=\mathbf{p}_1+\mathbf{p}_2$, where the action transfer direction is based on GM spectra. For PSI and ES, $+1$ denotes one unit of action received by the wave mode as a sink (red dot), and $-1$ denotes one unit of action sent by a mode as a source (green dot). Induced diffusion can reverse direction with sinks and sources (grey dots), which is determined by the spectral slopes to be discussed in \S 3.2. For all three mechanisms, the action transfer regarding the highest-frequency wave $\mathbf{p}_0$ is always opposite to those regarding $\mathbf{p}_1$ and $\mathbf{p}_2$, with energy conservation guaranteed by $\omega_0(\partial n_0/\partial t)=\omega_1(\partial n_1/\partial t) + \omega_2(\partial n_2/\partial t)$.}
 \label{fig_triad}
\end{figure}

The evaluation of IGW energy cascade for Garrett-Munk spectra based on \eqref{eq_KE} was first undertaken by McComas~\textit{et~al.} in a series of papers \citep{McComas1977a,McComas1981a,McComas1981b}. A major argument made in these works is that the collision integral in \eqref{eq_KE} is dominated by three types of nonlocal (i.e., scale-separated in either vertical wavenumber or frequency, or both) interactions, namely parametric subharmonic instability (PSI), elastic scattering (ES) and induced diffusion (ID) (see schematic illustrations in Figure~\ref{fig_triad}). 
\cite{McComas1981a} further argued that the Garrett-Munk spectrum is in equilibrium with respect to ES so that the downscale energy flux can be calculated from PSI and ID contributions. This simplification allowed an analytical formulation of the downscale energy flux, which laid the foundation of finescale parameterization that estimates the turbulent dissipation rate due to internal wave breaking. The Gregg–Henyey–Polzin (GHP) finescale parameterization yields \citep{Henyey1986, Gregg1989,Polzin1995,Polzin2014}
\begin{equation} \label{finescale}
P_\text{finescale} = C_0 \frac{fN^2\cosh^{-1}(N/f)}{f_0N_0^2\cosh^{-1}(N_0/f_0)}\hat{E}^2 \frac{3(R_{\omega}+1)}{4R_{\omega}} \sqrt{\frac{2}{R_{\omega}-1}}
\end{equation}
where $f_0=7.8361\times 10^{-5}$ s$^{-1}$ is the GM76-referenced Coriolis frequency corresponding to 32.5$^{\circ}$ Latitude, $N_0=3$ cph $= 5.2360\times 10^{-3}$ s$^{-1}$ is the GM76-referenced buoyancy frequency, $\hat{E}=0.1 \text{~cpm}/m_c$ is the nondimensional gradient spectral level with the critical vertical wavenumber $m_c$ evaluated through the integrated shear spectrum $\int_0^{m_c} 2m^2E_k(m)\,dm=2\pi N^2/10$, and $R_{\omega}$ is the shear-to-strain ratio. $C_0=8\times 10^{-10}$ W kg$^{-1}$ is a prefactor obtained by fitting microstructure measurements in the ocean. For GM76, $\hat{E}=1$ and $R_{\omega}=3$.
While \cite{McComas1981a} provides the correct functional form of \eqref{finescale}, their calculated value of $C=1.8 \times 10^{-9}$ W kg$^{-1}$ has a factor of 2.25 difference from $C_0$.

While McComas~\textit{et~al.}'s calculation provides an estimate of energy flux in the same order of finescale parameterization, the interaction mechanisms involved in the calculation suffer from physical \emph{inconsistencies} that have never been completely resolved. As summarized in \cite{Polzin2011}, at least two confusing paradoxes exist:
\begin{enumerate}[leftmargin=.6cm,labelsep=.1cm,label={(\alph*)}]
\item \noindent In frequency space, both dominant mechanisms of PSI and ID are believed to transfer energy toward low frequencies (i.e., backward cascade). This is not realistic for a balanced IGW continuum unless there is an energy injection at high frequency into the ocean, which is not known. Thus, there must exist a ``missing'' mechanism that moves energy to high frequencies to form a forward cascade. 
\item \noindent The GM76 wave action spectrum is independent of vertical wavenumber $m$ in the range of high $m$, which leads to a zero-flux state for ID (since diffusion requires gradient in $m$ at least at the leading order). This makes obscure McComas~\textit{et~al.}'s calculation where an artificial correction in ID has to be applied to enable cascade in the high-$m$ high-$\omega$ range of the spectra and raises questions on what the actual mechanism is for such cascade in this range.
\end{enumerate}

The paradoxes (a) and (b) have been partly addressed by \cite{Dematteis2021} and \cite{Dematteis2022}, mainly for a modified GM76 spectrum that serves as a stationary solution to \eqref{eq_KE} [action spectrum $n(k,m)\sim k^{-3.69}m^0$ instead of the standard GM76 $n(k,m)\sim k^{-4}m^0$]. In these works, it was necessary to consider the non-rotating condition $f=0$ such that \eqref{eq_KE} becomes scale-invariant and yields a power-law solution. For this modified power-law spectrum, it was identified that ID provides a non-zero and frequency-forward flux by considering the complete diffusion tensor (i.e., including the off-diagonal elements) and that local interactions (which had been ignored in McComas~\textit{et~al.}) play a major role in the downscale energy cascade. By applying a combined analytical and numerical approach, the authors explicitly calculated the downscale energy flux, which is within a factor of 2 compared to the prediction by finescale parameterization \eqref{finescale}.

In spite of the significant progress achieved in \cite{Dematteis2021} and \cite{Dematteis2022}, the paradoxes (a) and (b), along with a convincing evaluation of downscale energy flux in quantitative consistency with \eqref{finescale} in the WKE framework, have never been addressed for the original problem of the GM76 spectrum. This is an important task considering that the GM76 spectrum, although not a stationary state of \eqref{eq_KE} as understood now, is still largely considered as a standard model for realistic IGW spectra among most physical oceanographers. We undertake this task leveraging the fast rise of computational power that has enabled a direct numerical evaluation of the complete spectral energy transfer based on \eqref{eq_KE}. Since our approach is purely numerical and we do not seek a scale-invariant field, we are able to naturally incorporate a finite value of $f$ that was not treated in \cite{Dematteis2021} and \cite{Dematteis2022}. 
Our results show that the energy flux across the critical vertical scale of 10 m (generally considered as the scale where dissipation starts to take over linear wave dynamics) is approximately $1.5 \times 10^{-9}$ W kg$^{-1}$, with a factor of 1.5 greater than the finescale formula \eqref{finescale}. 
We further decompose the cascade into different mechanisms and show that the downscale flux is mainly provided by PSI and local interactions, with ID contributing almost zero flux. This is in sharp contrast to McComas~\textit{et~al.}'s calculation (flux based on PSI plus ID) and addresses the paradox (b). In addition, we find that there exists a clear frequency-forward cascade, supplied mainly by local interactions and ES, which addresses paradox (a). The role of ES, which was previously hypothesized to have no effect on energy cascade in GM76, is now revealed because of the adoption of finite Coriolis frequency $f$.

\section{Numerical Method}

For numerical evaluation of spectral energy transfer, we use the WKE derived in \cite{Lvov2001, Lvov2004b} with detailed formulation of the interaction coefficient $V$ provided in \cite{Lvov2010, Lvov2012} and \cite{Pan2020}. This WKE is derived from a Hamiltonian formulation of the dynamical equation of IGWs, in which the isopycnal vertical coordinate has to be used. The isopycnal vertical wavenumber, $m_i$, is converted from its Eulerian counterpart by $m_i=(g/\rho N^2)\,m$. Hereafter, we use notation $m$ throughout the paper for convenience, which represents isopycnal $m_i$ in the formulation of WKE [e.g., \eqref{reduceKE} and the appendices] and Eulerian $m$ in presenting the results in \S 3.

The WKE, in the form of \eqref{eq_KE}, involves a collision integral over six dimensions in $\mathbf{p}_1$ and $\mathbf{p}_2$. One can reduce the dimension of integration by integrating only on the resonant manifold defined by the delta functions. Since GM76 spectrum is horizontally isotropic, it is convenient to first integrate out the horizontal angle dependence, then reduce the integration by applying the delta functions. With detailed formulation provided in Appendix A, this procedure leads to an integration over only two dimensions, namely magnitudes of horizontal wavenumbers $k_1$ and $k_2$
\begin{equation} \label{reduceKE}
\frac{\partial n(k,m)}{\partial t} = \int_0\displaylimits^{+\infty} \int\displaylimits_0^{+\infty} \left[\frac{h^+(m^{*+}_1)}{|{g^+}'(m^{*+}_1)|}-2\frac{h^-(m^{*-}_1)}{|{g^-}'(m^{*-}_1)|}\right] \,dk_1\,dk_2,
\end{equation}
where functions $h^+$, $h^-$, ${g^+}'$, ${g^-}'$ (which additionally depend on $k,m,k_1,k_2$) and the roots $m^{*+}$ and $m^{*-}$ are defined in Appendix A, $k=\sqrt{k_x^2+k_y^2}\in \mathbb{R}^+$, and $m\in \mathbb{R}$ (taking both positive and negative values). The numerical integration of \eqref{reduceKE} is rather straightforward, but care has to be taken in terms of the root finding for $m^{*+}$ and $m^{*-}$ with details discussed in Appendix B. Our numerical code, implemented in FORTRAN with Message Passing Interface (MPI) for parallel computation, is made available on GitHub at https://github.com/yue-cynthia-wu.

Our numerical approach, in principle, shares some level of similarity to ``Method 3'' in \cite{Eden2019b} regarding the evaluation of the collision integral \eqref{reduceKE} on the resonant manifold, but the latter is implemented for a different version of WKE. Additionally, our procedure (of using cylindrical coordinates and integrating out the horizontal angle dependence) does enforce horizontal isotropy of the IGW spectrum. This feature is beneficial for our planned subsequent work (beyond this paper) to integrate the WKE in time while exactly maintaining the horizontal isotropy as was done in \cite{Olbers2020}. In addition, \cite{Eden2019b} employed other methods using broadened delta functions to compute the collision integral, but the results are more noisy and do not show clear advantage. Indeed, as understood recently in both pure mathematical derivation \cite[e.g.][]{Deng2023} and numerical studies \cite[e.g.][]{Hrabski2020,Hrabski2022}, the WKE should be considered as a result of maintaining sufficient quasi-resonances from the dynamical equation, so using broadening in the delta function is somewhat redundant. Such broadened delta functions, on the other hand, might be physically meaningful if finite size effect is important, as in situations described in \cite{Pan2017a}. 

\begin{figure}
 \centerline{\includegraphics[width=\textwidth]{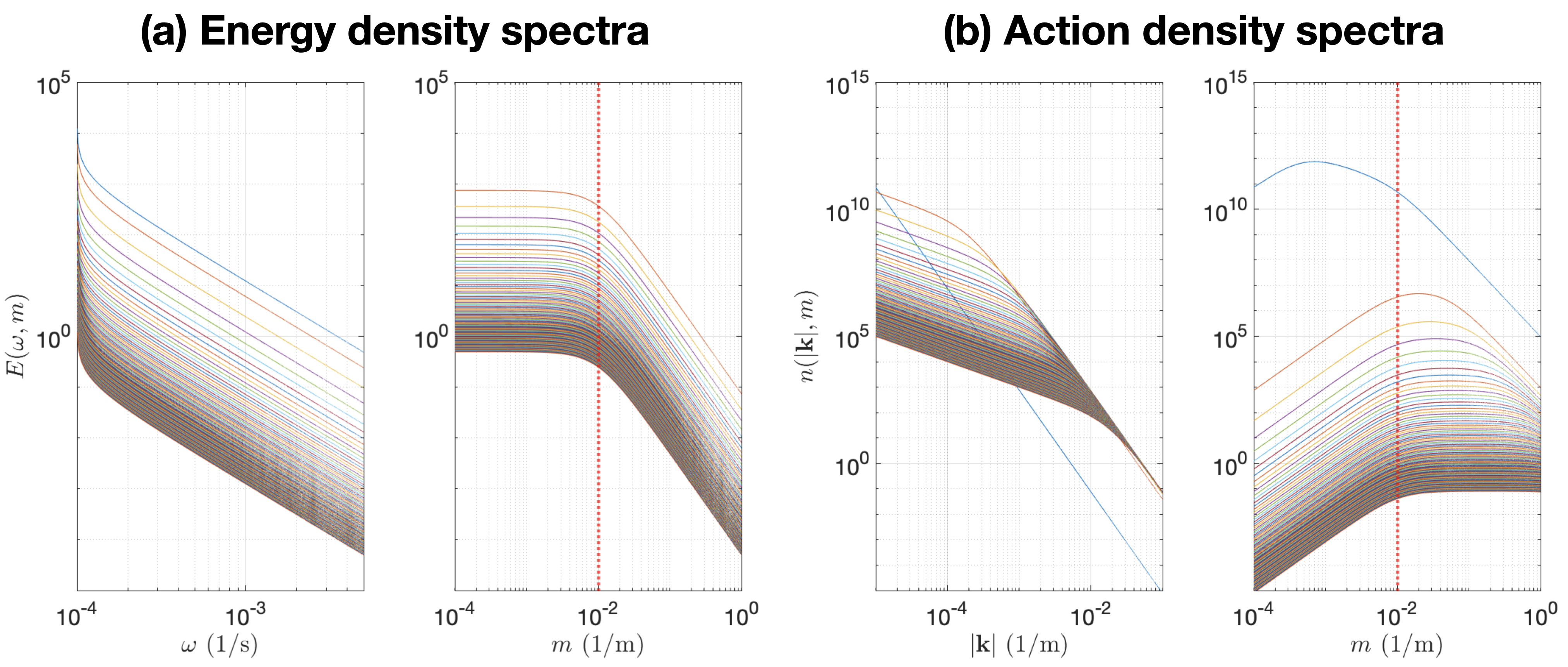}}
 \caption{GM76 model of the (left) energy and (right) action density spectra, with $E(\omega,m) \sim \omega^{-2} m^{-2}$ and $n(k,m) \sim k^{-4} m^{0}$ in the high-frequency, high-wavenumber regime. Curves in each figure represent the spectrum with one variable taking fixed values. Red vertical lines denote the GM76-referenced vertical wavenumber $m^*=10^{-2}$ m$^{-1}$.}
 \label{fig_GM}
\end{figure}

Despite the observed variability in the spectral forms of the realistic IGW fields in different seasons and at different geographical locations, we start with the GM76 model which is one of many realistic possible spectra 
\begin{equation}
E(\omega,m)=E_0 A(m/m^*)B(\omega),
\end{equation}
where $m^*=10^{-2}$ m$^{-1}$ is a reference vertical wavenumber, and the functions
\[A(m/m^*)\sim m^{*-1}[1+(m/m^*)^2]^{-1},\] \[B(\omega)\sim \omega^{-1}(\omega^2-f^2)^{-1/2},\]
are normalized to be integrated to unit [i.e., $\int_0^\infty A(m/m^*) d(m/m^*)=1$ and $\int_f^N B(\omega) d\omega=1$ with $f=10^{-4}$ s$^{-1}$ and $N=5\times10^{-3}$ s$^{-1}$] so that the total (integrated) energy level of GM76 equals $E_0=3\times10^{-3}$ m$^2$ s$^{-2}$. 
The action density spectra, which is needed in evaluating \eqref{reduceKE}, can be calculated by $n(k,m)=E(k,m)/\omega=E(\omega,m)(\partial \omega/\partial k)/(2\pi \omega k)$ considering horizontal isotropy. In the high-$\omega$ high-$m$ regime of the spectrum, we have $E(\omega,m) \sim \omega^{-2} m^{-2}$ and $n(k,m) \sim k^{-4} m^{0}$ (see Figure~\ref{fig_GM}).

In order to set our computation for a physical problem that reflects the size of the real ocean, we consider a domain with a horizontal circular radius of 42.4 km and a vertical extent of 2.1 km. To evaluate \eqref{reduceKE}, we use 1080$\times$1080 grids of uniform spacing in both $k$ and $m$, with the smallest resolved scales to be 40 m and 2 m in the two directions, i.e., $k \in [1.5\times10^{-4}, 1.6\times10^{-1}]$ m$^{-1}$ and $m \in [3.0\times10^{-3}, 3.2]$ m$^{-1}$.
We use the nonhydrostatic dispersion relation $\omega^2=(N^2k^2+f^2m^2)/(k^2+m^2)$ which bounds the IGW frequency between $f$ and $N$, while the hydrostatic version leads to (non-physical) super-buoyancy IGWs in large spectral area of the selected $(k,m)$-domain where $k/m$ is large\footnote{Since the interaction coefficients in the WKE used in this paper are derived under hydrostatic approximation, the results at regions of large $k/m$ should be considered as an approximation. \cite{Olbers1974, Olbers1976} and \cite{Muller1975a} provided a nonhydrostatic version of the WKE for IGWs.}. With above setting, our numerical calculations are performed on the Great Lakes clusters at University of Michigan with 2 nodes of 72 CPUs, and the simulation takes 6--8 hours to calculate all $O(10^{12})$ triad interactions. 

\section{Results}
\subsection{Energy transfer in spectral space}

\begin{figure}
 \centerline{\includegraphics[width=.8\textwidth]{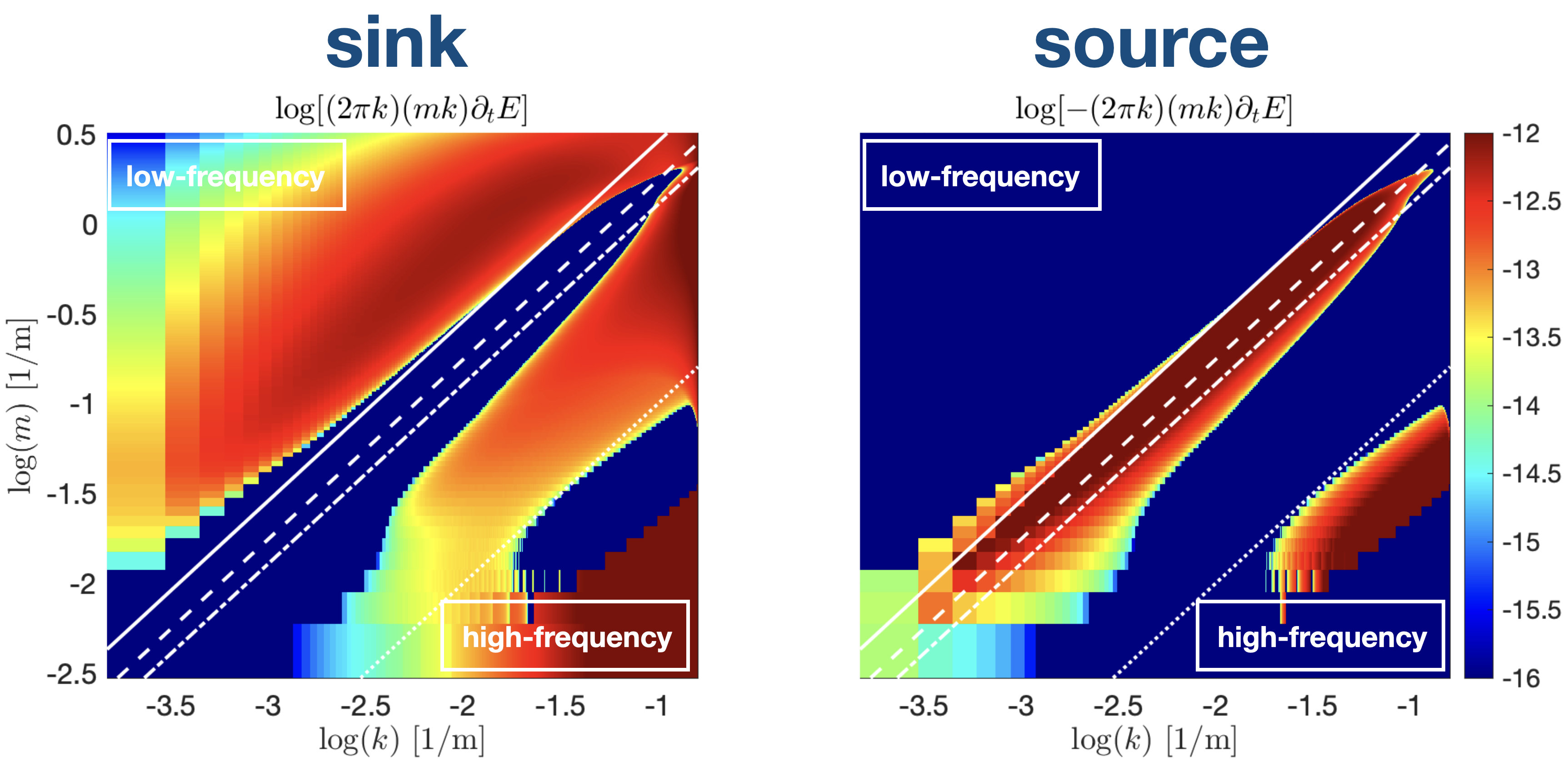}}
 \caption{Energy transfer $\log[(2\pi k)(mk)\partial E/\partial t]$ for GM76 with sinks ($\partial E/\partial t>0$) on the left and sources ($\partial E/\partial t<0$) on the right. We only plot the range with positive $m$ since the spectrum at negative $m$ is completely symmetric. The white solid, dashed, dash-dotted and dotted lines along the diagonal denote frequencies $2f$, $3f$, $4f$ and $35f (=0.7N)$, respectively.}
 \label{fig_Et}
\end{figure}

The energy transfer in spectral space $\partial E(k,m)/\partial t$ is calculated by multiplying $\partial n(k,m)/\partial t$ with $\omega$, where $\partial n(k,m)/\partial t$ is obtained by numerically evaluating the WKE \eqref{reduceKE}. In Figure~\ref{fig_Et}, we plot $(2\pi k)(mk)\,\partial E(k,m)/\partial t$, with the factor $2\pi k$ accounting for horizontal azimuth integration and $mk$ accounting for the plot in log-log axis. More precisely, with these prefactors, the total $\partial E/\partial t$ can be conveniently computed by integrating the values over the area in Figure~\ref{fig_Et}, i.e., $\partial E/\partial t = \iint (2\pi k)(mk)\partial E(k,m)/\partial t \,d[\log (m)]\,d[\log(k)]$ (we will later show in Figure~\ref{fig_P} that our simulation conserves the total energy so that $\partial E/\partial t=0$). This plotting technique is also used in \cite{Eden2019a,Eden2019b}, which facilitates an unbiased visualization of energy transfer. We further split Figure~\ref{fig_Et} into two panels, showing respectively the source [with $\partial E(k,m)/\partial t<0$, providing energy] and sink [with $\partial E(k,m)/\partial t>0$, receiving energy] regions. 
Here, the terminologies ``sink" and ``source" are inherited from \cite{Eden2019b} and are used to indicate the direction of energy cascade. If a stationary spectrum is assumed, they can be physically related to the generation and dissipation mechanism that has to balance the spectral energy transfer.
We see that energy is transferred from waves in the frequency band $[2f,4f]$ to both lower and higher frequencies. The component of forward frequency cascade is also seen in the results of \cite{Eden2019b}. We shall further investigate the mechanisms leading to this cascade in \S 3.2, addressing paradox (a) in McComas~\textit{et~al.}'s theory.

\begin{figure}
 \centerline{\includegraphics[width=.6\textwidth]{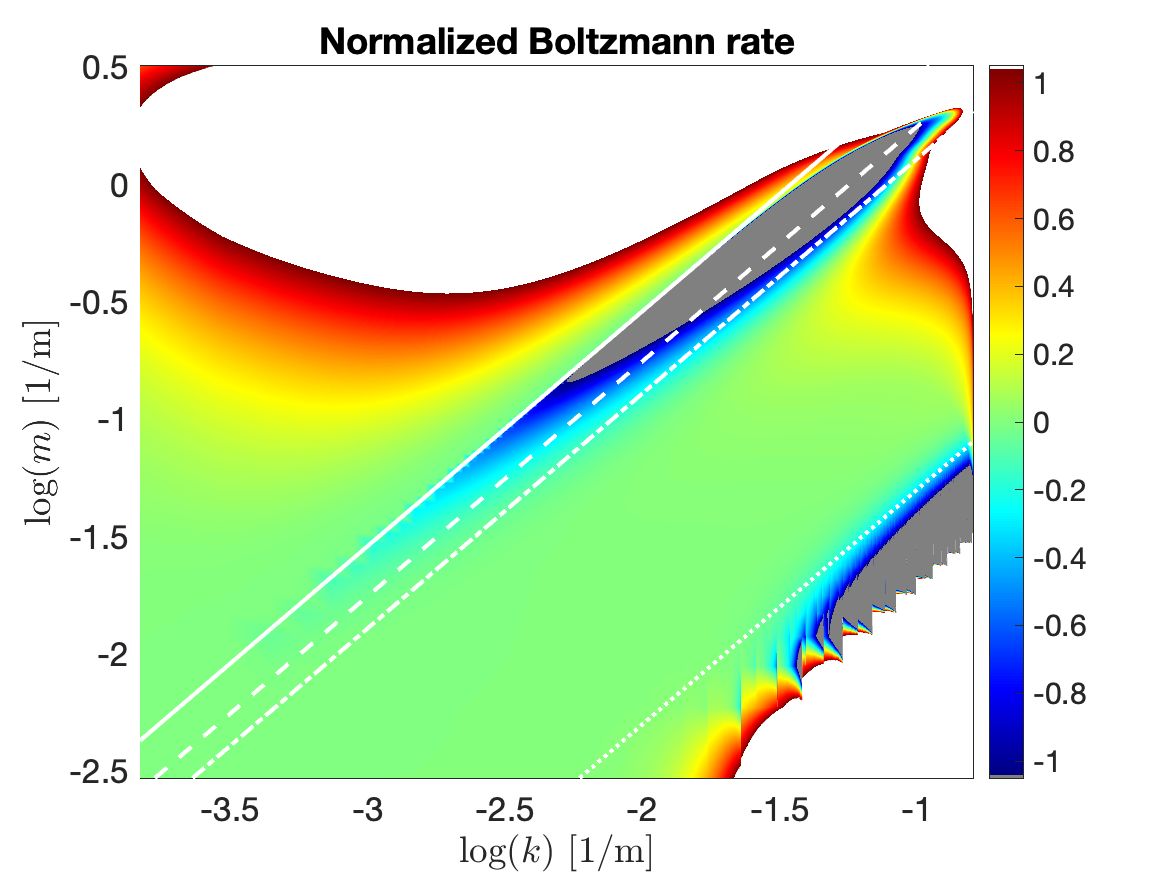}}
 \caption{Normalized Boltzmann rate $\varepsilon_\mathbf{p}=2\pi (\partial n_\mathbf{p}/\partial t)/(\omega_\mathbf{p} n_\mathbf{p})$ for GM76 in $(k,m)$-domain. 
 The white solid, dashed, dash-dotted and dotted lines along the diagonal denote frequencies $2f$, $3f$, $4f$ and $45f (=0.9N)$, respectively.
 Regions where $m>0.1$ m$^{-1}$ and/or $\omega>0.9N$ indicate violation of the weak nonlinearity assumption in wave turbulent theory, including regions of $\varepsilon_\mathbf{p}>1$ (white) and $\varepsilon_\mathbf{p}<-1$ (grey) with obvious violation.}
 \label{fig_Bo}
\end{figure}

\begin{figure}
 \centerline{\includegraphics[width=.6\textwidth]{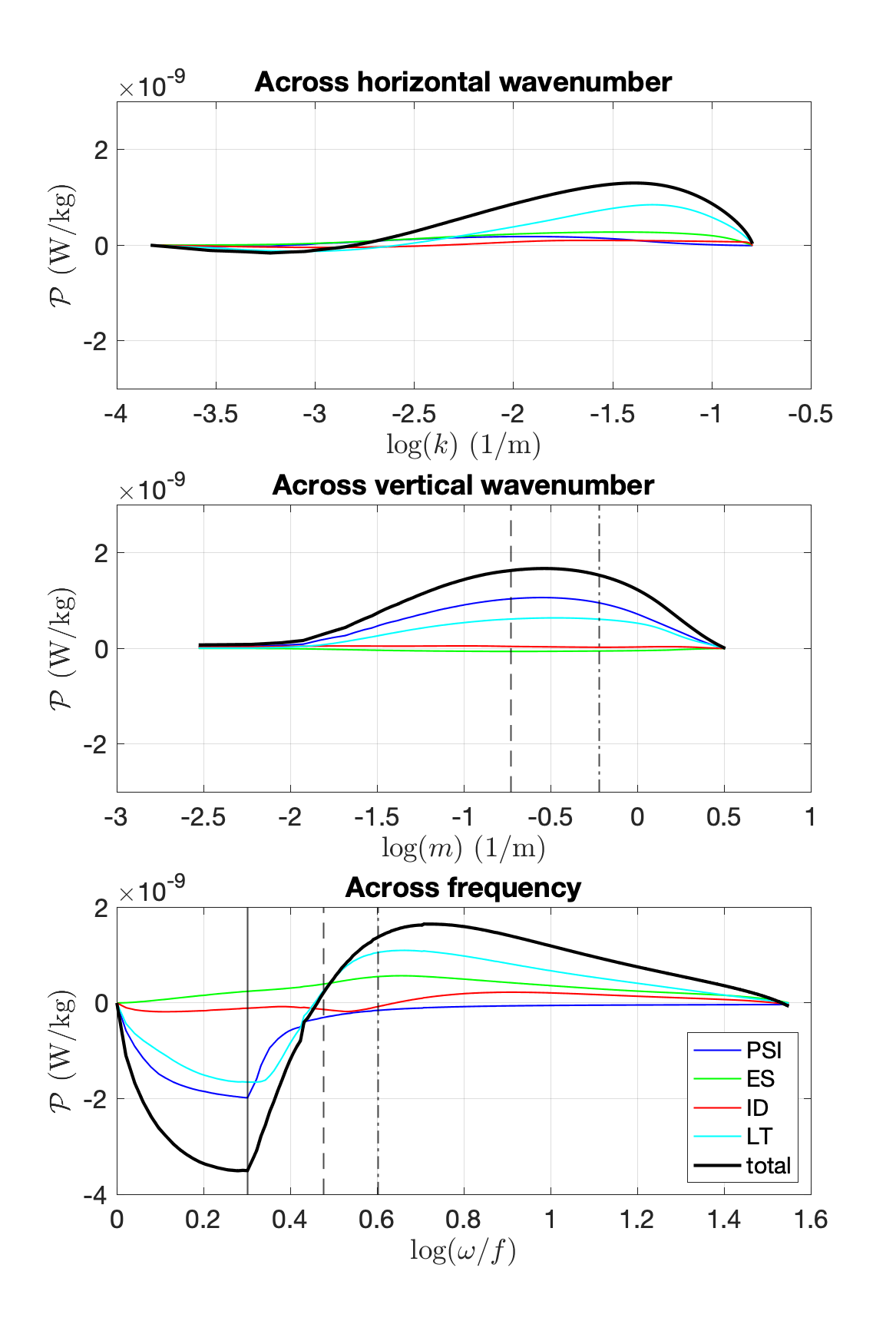}}
 \caption{Energy fluxes due to nonlinear wave-wave interactions within the selected scale limits across (a) horizontal wavenumbers, (b) vertical wavenumber, and (c) frequency for GM76. In (b), dashed and dash-dotted lines denote $m_\text{cutoff}=0.2$ m$^{-1}$ (corresponding to the smallest scale that 90\% of the waves with $|\varepsilon_\mathbf{p}|<0.2$; Figure~\ref{fig_Bo}) and the critical vertical wavenumber $m_c=0.6$ m$^{-1}$, respectively. In (c), solid, dashed and dash-dotted lines denote $\omega=2f,3f$ and $4f$, respectively. Colored curves denote PSI, ES, ID and local interactions (LT) using selection criteria defined in \S 3.2.}
 \label{fig_P}
\end{figure}

It is important to verify that the WKE under the above setting stays in the weakly nonlinear regime and provides a valid approximation of the dynamics. In particular, one may concern about the rapid modal evolution in the high-wavenumber high-frequency regime, as first pointed out in \cite{Holloway1978}, which may violate condition \eqref{eq_Bo} regarding the normalized Boltzmann rate $\varepsilon_\mathbf{p}$. 
For this purpose, we check the values of $\varepsilon_\mathbf{p}$ in the selected scale limits. As shown in Figure~\ref{fig_Bo}, there indeed exist large spectral regions where $m>0.1$ m$^{-1}$ and/or $\omega>0.9N$ with $|\varepsilon_\mathbf{p}| \sim O(1)$, indicating the failure of WKE in the regime of high wavenumbers and/or high frequencies. These regions with $|\varepsilon_\mathbf{p}|$ not much less than 1 will be treated with caution in the subsequent discussion of energy fluxes.

We further define the energy fluxes across $k_0$, $m_0$ and $\omega_0$, respectively, based on energy conservation in finite domain
\begin{align} \label{eq_Pk}
&\mathcal{P}^k(k_0)=-\int\displaylimits_0^{m_{\max}} \int\displaylimits_0^{k_0} 4\pi k \frac{\partial E(k,m)}{\partial t} \,dk \,dm,\\
\label{eq_Pm} &\mathcal{P}^m(m_0)=-\int\displaylimits_0^{m_0} \int\displaylimits_0^{k_{\max}} 4\pi k \frac{\partial E(k,m)}{\partial t} \,dk \,dm,\\
\label{eq_Po} &\mathcal{P}^\omega(\omega_0)=-\int\displaylimits_{0}^{m_{\max}} \int\displaylimits_0^{k_\text{max}} 4\pi k \frac{\partial E(k,m)}{\partial t} \, \mathbbm{1}_{\omega\le\omega_0}\,dk \,dm, 
\end{align}
where $k_{\max}=0.016$ m$^{-1}$, $m_{\max}=0.32$ m$^{-1}$, and $\mathbbm{1}$ an indicator function. 
Equation \eqref{eq_Pk}--\eqref{eq_Po} are energy fluxes only due to nonlinear wave-triad interactions within the selected scale limits, which do not include fluxes entering the IGW field from the largescale end by generation nor draining from the smallscale end by dissipation.
The evaluations here are based on the conservation of total energy $\partial E/\partial t$ by the WKE. In the prefactors $4\pi k$, $2\pi k$ comes from the integration over horizontal azimuth and $2$ accounts for the vertical symmetry over $\pm m$. The energy flux in all three directions are plotted in Figure~\ref{fig_P} (black curves). We see that $\mathcal{P}^\alpha (\alpha_\text{max}) \approx 0$ with $\alpha=k, m$ and $\omega$, indicating energy conservation. 
We remark here that energy conservation is only approximately achieved by our numerical algorithm since the roots $m^{*+}$ and $m^{*-}$ in \eqref{reduceKE} found by the root-finding algorithm (Appendix B) do not exactly lie on the discrete $m$-grid points, which breaks the symmetry when looping over three wavenumber vectors in a triad. This mainly affects the high-frequency regime of the spectrum (Figure~\ref{fig_Et} with strong sink and source where $\omega>0.7N$)\footnote{The region where $\omega \sim N$ is problematic because of two reasons: (1) if $m$ is not much greater than $k$, the hydrostatic approximation in the WKE breaks down, making the evaluation of $\partial n/\partial t$ (and $\partial E/\partial t$) untrustworthy, 
and (2) if $m \in \mathbb{R}^+$ is small, the root-finding for the reduction interactions ($m^*_{1,\text{right}} \in (0,m)$) deviates from its true value due to limited resolution between 0 and $m$ (Appendix B and Figure~8).}. 
Since the hydrostatic approximation is also violated in this regime, we discard the contribution of triads with frequencies greater than the cutoff frequency $\omega_\text{cutoff}=0.7N$ in the calculation of energy fluxes.

The fluxes defined in Equation \eqref{eq_Pk}--\eqref{eq_Po} are energy transfer only due to nonlinear wave-wave interactions within the selected scale limits, and the values have to be zero at the boundaries (e.g., $k_\text{min}$ and $k_\text{max}$, etc.) due to energy conservation. It is clear from Figure~\ref{fig_P} that the GM76 spectrum, as expected, does not yield a constant energy flux in any spectral direction. While the flux across frequency is bi-directional [Figure~\ref{fig_Et} and \ref{fig_P}(c)], the fluxes across horizontal and vertical wavenumbers are mainly in the forward direction (toward small scales). 

To evaluate the downscale energy flux $\mathcal{P}_d$ that provides energy for small-scale dissipation and mixing, i.e., to evaluate the finescale parameterization, we may follow two approaches. The first approach is to consider $\mathcal{P}_d=\mathcal{P}^m(m_c)$ with the critical vertical wavenumber $m_c$ evaluated through $\int_0^{m_c} 2m^2E_k(m)\,dm=2\pi N^2/10$ \citep{Polzin2014}. 
This approximation encapsulates the energy escaping the internal-wave field at the critical wavenumber $m_c\approx 0.6$ m$^{-1}$ past which internal waves become unstable to shear instability.
The first approach gives $\mathcal{P}_d=\mathcal{P}^m(m_c)=1.5 \times 10^{-9}$ W kg$^{-1}$, with a factor of 1.5 greater than $P_\text{finscale}=1.0 \times 10^{-9}$ W kg$^{-1}$ from \eqref{finescale} (equation rescaled for our values of $f$ and $N$).
However, as seen in Figure~\ref{fig_Bo}, the Boltzmann rate close to $m_c$ contains large regions of values that are not much less than 1, making the validity of WKE questionable. 
The second approach is to instead define a cutoff vertical wavenumber $m_\text{cutoff}\approx 0.2$ m$^{-1}$, below which only 10\% of the
computed waves violate the weak nonlinearity assumption with $|\varepsilon_\mathbf{p}|>0.2$. The second approach gives $\mathcal{P}_d=\mathcal{P}^m(m_\text{cutoff})=1.6 \times 10^{-9}$ W kg$^{-1}$. This approximation yields an upper bound of energy available for dissipation while (almost) free of uncertainties associated with the first approach. 
We should acknowledge that there exists a gap between the (vertical) length scales where WKE breaks down and the scale of 10 meter in the vertical.
Fortunately, the energy-flux curve between the two scales (corresponding to $m_c\approx 0.6$ m$^{-1}$ and $m_\text{cutoff}\approx 0.2$ m$^{-1}$) is relatively insensitive to the vertical wavenumber (Figure~\ref{fig_P}b), making our estimate quite robust.

\subsection{Contributions of PSI, ES and ID triads}
In this section, we discuss the decomposition of energy transfer into contributions from different mechanisms, namely nonlocal interactions of PSI, ES and ID, and local interactions. The nonlocal interactions exhibit scale separation in frequency or wavenumber, or both, based on which our classification method is designed. 
For a resonant triad, we rank the frequencies from high to low as $(\omega^H,\omega^M,\omega^L)$ and the magnitude of vertical wavenumbers as $(|m^H|,|m^M|,|m^L|)$.
We can then naturally classify nonlocal interactions according to some threshold values of the two metrics as follows:
\begin{itemize}[leftmargin=.6cm,labelsep=.1cm]
\item PSI: \text{~} $|m^M|/|m^L|>\eta$, \text{~} $1/2 \le \omega^M/\omega^H <1/2+\epsilon/2$
\item ES: \text{~~~} $\omega^M/\omega^L>\xi$, \text{~~~~~} $1/2 \le |m^M|/|m^H| <1/2+\alpha/2$
\item ID: \text{~~~} $\omega^M/\omega^L>\xi$, \text{~~~~~} $|m^M|/|m^L|>\eta$
\end{itemize}

The above criterion characterizes PSI as scale-separated in $m$ and halving in $\omega$, ES as scale-separated in $\omega$ and halving in $m$, and ID as scale-separated in both $m$ and $\omega$. In practice, we use $\xi=\eta=2$ and $\epsilon=\alpha=0.1$ for results below, but we note that the major conclusions hold for a reasonable range of parameters selected.
We also remark that the above choices of $\xi$ and $\eta$ are ``conservative'' for local interactions, in the sense that some interactions with moderate $\xi$ and $\eta$ (say slightly greater than 2) are classified as nonlocal, instead of local, interactions. This is not a drawback for this paper, as we shall show that even for this ``conservtive'' choice of local interactions, their role in energy cascade is significant, in sharp contrast to McComas~\textit{et~al.}'s early conjecture. 
In the following sections, we discuss the roles of each mechanism in spectral energy transfer.

\subsubsection*{\textbf{The PSI mechanism}}
Parametric subharmonic instability represents the decay of a low-wavenumber parent wave into two nearly identical high-wavenumber child waves with half frequencies. 
One unit of action of the parent wave $\mathbf{p}_0$ is transferred into two units of action of the child waves $\mathbf{p}_1$ and $\mathbf{p}_2$ (see Figure~\ref{fig_triad}a).

\begin{figure}
 \centerline{\includegraphics[width=\textwidth]{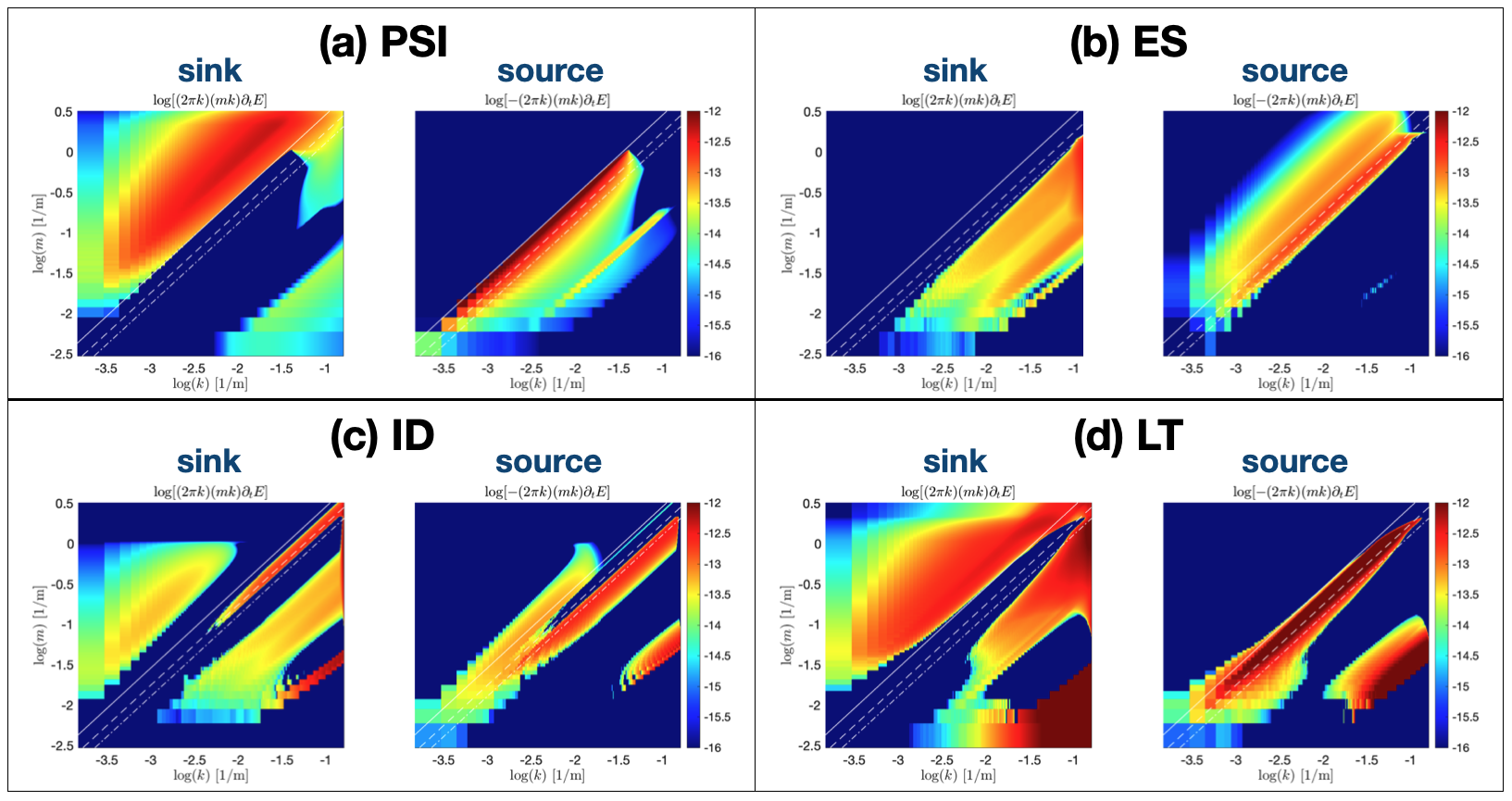}}
 \caption{As Figure~\ref{fig_Et} but for different mechanisms.}
 \label{fig_Et_sep}
\end{figure}

Using the criterion defined above, we compute the spectral energy transfer due to PSI, with result shown in Figure~\ref{fig_Et_sep}a. In terms of energy cascade in frequency, we see that PSI contributes predominantly to the backward cascade, namely moving energy from frequency band $[2f,4f]$ (source) to $[f,2f]$ (sink). This frequency cascade is accompanied by a strong forward cascade in vertical wavenumbers, which is also clear from the figure. The physical picture of PSI revealed here is consistent with the existing understanding that PSI is effective in tidal damping (i.e., extracting energy above $2f$) and that PSI contributes significantly to downscale energy cascade in the IGW continuum \citep{McComas1977a}.

\subsubsection*{\textbf{The ES mechanism}}
The energy transfer due to ES is plotted in Figure~\ref{fig_Et_sep}b, which shows a clear forward cascade in frequency. This was not understood by previous theory of McComas~\textit{et~al.}, which instead postulated that ES can be neglected for energy transfer in any vertically symmetric IGW spectrum. In fact, the previous postulation to neglect ES' contribution is a bit surprising given that the dynamics of ES is similar to a diffusion process that can be understood in analogy to ID (the previous researchers do have a better understanding of ID as will be discussed later in the paper). Consider an ES triad $\mathbf{p}_0=\mathbf{p}_1+\mathbf{p}_2$ (Figure~\ref{fig_triad}b), where $\mathbf{p}_2$ is the near-inertial mode, and $\mathbf{p}_0$ and $\mathbf{p}_1$ the high-frequency modes with $\omega_0 \approx \omega_1 + f$. Given the fact that the action spectra are red with respect to $\omega$, i.e., most action contained in low frequencies, it is reasonable to set $n_2\gg n_0,n_1$ and $n_0<n_1$. According to WKE, we then have
\begin{align}
&\partial n_0/\partial t= C f_{012} = C(n_1n_2 -n_0n_1-n_0n_2) \approx Cn_2(n_1-n_0) >0,\nonumber\\
&\partial n_1/\partial t=-C f_{012} =-C(n_1n_2 -n_0n_1-n_0n_2) \approx -Cn_2(n_1-n_0) <0,\nonumber\\
&\partial n_2/\partial t=-C f_{021} =-C(n_1n_2 -n_0n_1-n_0n_2) \approx -Cn_2(n_1-n_0) <0,\nonumber
\end{align}
where $C=4\pi|V(\mathbf{p}_0,\mathbf{p}_1,\mathbf{p}_2)|^2$ is a constant for this triad. The sign of $\partial n/\partial t$ indicates $\mathbf{p}_0$ is a sink while $\mathbf{p}_1$ and $\mathbf{p}_2$ are sources. Consumption of one unit of action of $\mathbf{p}_2$ combined with one unit of action of $\mathbf{p}_1$ results in generation of one unit of action of $\mathbf{p}_0$. This process can be equivalently described as diffusion from $\mathbf{p}_1$ to $\mathbf{p}_0$ (i.e., toward higher frequency) which in the meanwhile extracts energy from $\mathbf{p}_2$. 

It is clear from the above analysis that a finite value of $f$ is critical to enable the forward cascade in frequency, which we indeed incorporate in our calculation. The assumption of $f=0$ used in previous research (either for convenience or for obtaining a scale-invariant WKE as in \citealt{Dematteis2021,Dematteis2022}) is perhaps one reason leading to the neglect of ES in energy transfer. With the above analysis, we can conclude that forward frequency cascade by ES shown in Figure~\ref{fig_Et_sep}b should essentially be expected, and we reach consistency in theory and numerical results. 

\subsubsection*{\textbf{The ID mechanism}}
The energy flux due to ID is plotted in Figure~\ref{fig_Et_sep}c, which shows a very weak transfer compared to those from other interaction mechanisms. One could further expect that ID contributes insignificantly to the downscale energy cascade for GM76 spectrum. This is in strong disagreement with results in \cite{McComas1981a, McComas1981b} that ID contributes nearly 30\% of the total downscale energy cascade. The result from McComas~\textit{et al.} roots from the hypothesis that GM76 spectrum yields a constant downscale energy flux which relies on a logarithmic correction to the ID flux. It is clear from our analysis (Figure~\ref{fig_P}) that the constant flux hypothesis is incorrect, and thus the logarithmic correction has no meaningful ground.

The ID mechanism for GM76 or more general IGW spectra can be conveniently understood from a diffusion equation in the high-$m$ high-$\omega$ regime: $\partial n/\partial t=\partial /\partial p_i (\mathcal{D}_{ij}\partial n/\partial p_j)$ with $p_i=(k_x, k_y, m)$ for $i=(1,2,3)$ and $\mathcal{D}_{ij}$ as the diffusion coefficient matrix. This equation, including the detailed formulation of $\mathcal{D}_{ij}$, can be derived by taking dominant terms in the WKE or from a WKB approximation of the dynamic equation, which was first done in \cite{McComas1981b} and re-derived by \cite{Lvov2022} for IGWs [also see derivations in other physical contexts such as MHD turbulence \cite[][]{Nazarenko2001}, Rossby waves \cite[][]{Connaughton2015} and surface gravity waves \cite[][]{Korotkevich2023}]. With $D_{33}$ being the dominant element in $\mathcal{D}_{ij}$ for IGWs, the leading order effect of the diffusion takes place in the vertical wavenumber direction and can be approximated by a one-dimensional diffusion equation $\partial n/\partial t=\partial /\partial m (\mathcal{D}_{33}\partial n/\partial m)$. Along with this approximation, one can see that in the high-$m$ high-$\omega$ regime the direction of diffusion is fully determined by the dependence of wave action spectrum on $m$, i.e., action diffuses down gradient in vertical wavenumber direction, according to $n(k_0, m)$ at a given $k_0$.

The physical picture of ID revealed above through the diffusion equation can be alternatively explained directly via WKE using a similar argument as our previous one for ES. Take an ID triad $\mathbf{p}_0=\mathbf{p}_1+\mathbf{p}_2$ (Figure~\ref{fig_triad}c) with $\mathbf{p}_1$ the low-$m$ low-$\omega$ mode and consider $n_1\gg n_0,n_2$, we obtain from WKE that
\begin{align}
&\partial n_0/\partial t= C f_{012}= C(n_1n_2 -n_0n_1-n_0n_2) \approx C n_1(n_2-n_0),\nonumber\\
&\partial n_1/\partial t=-C f_{012}=-C(n_1n_2 +n_0n_1+n_0n_2) \approx -C n_1(n_2-n_0),\nonumber\\
&\partial n_2/\partial t=-C f_{021}=-C(n_1n_2 +n_0n_1+n_0n_2) \approx -C n_1(n_2-n_0).\nonumber
\end{align}
We see that the exchange of action between the two high-$m$ high-$\omega$ modes, $\mathbf{p}_2$ and $\mathbf{p}_0$, depends on the relative magnitudes between $n_2$ and $n_0$. The one that is greater between $n_2$ and $n_0$ serves as the source of the diffusion and the other as the sink. If we further assume that $\mathbf{p}_0$ and $\mathbf{p}_2$ have the same horizontal wavenumber $k_0$, i.e., $\mathbf{p}_1$ is (almost) vertical, then the diffusion direction is again determined by the dependence of $n(k_0, m)$ on $m$.

Now let us consider a general power-law spectrum in the high-$m$ high-$\omega$ regime: $E(\omega,m)\sim \omega^{-2-p} m^{-2-p-q}$, equivalent to $n(k,m)\sim k^{-4-p} m^{-q}$. Based on the above analysis (either from the diffusion equation or WKE), the leading-order diffusion direction is controlled only by the parameter $q$. For the GM76 spectrum with $p=q=0$, it is expected that the leading-order diffusion vanishes, i.e., the GM76 spectrum is indeed approximately a zero-flux state for ID. The energy transfer in Figure~\ref{fig_Et_sep}c, in fact, mainly comes from the (off-diagonal) sub-elements in $\mathcal{D}_{ij}$ (other than $\mathcal{D}_{33}$). The effects of the sub-diffusion is analyzed analytically in \cite{Dematteis2022} for the scale-invariant case. For our case with finite $f$ (which breaks the scale invariance), an analytical treatment is generally much difficult. Nevertheless, our direct numerical calculation shows that the transfer generated by the sub-diffusion is weak compared to other interaction mechanisms.

\begin{figure}
 \centerline{\includegraphics[width=.8\textwidth]{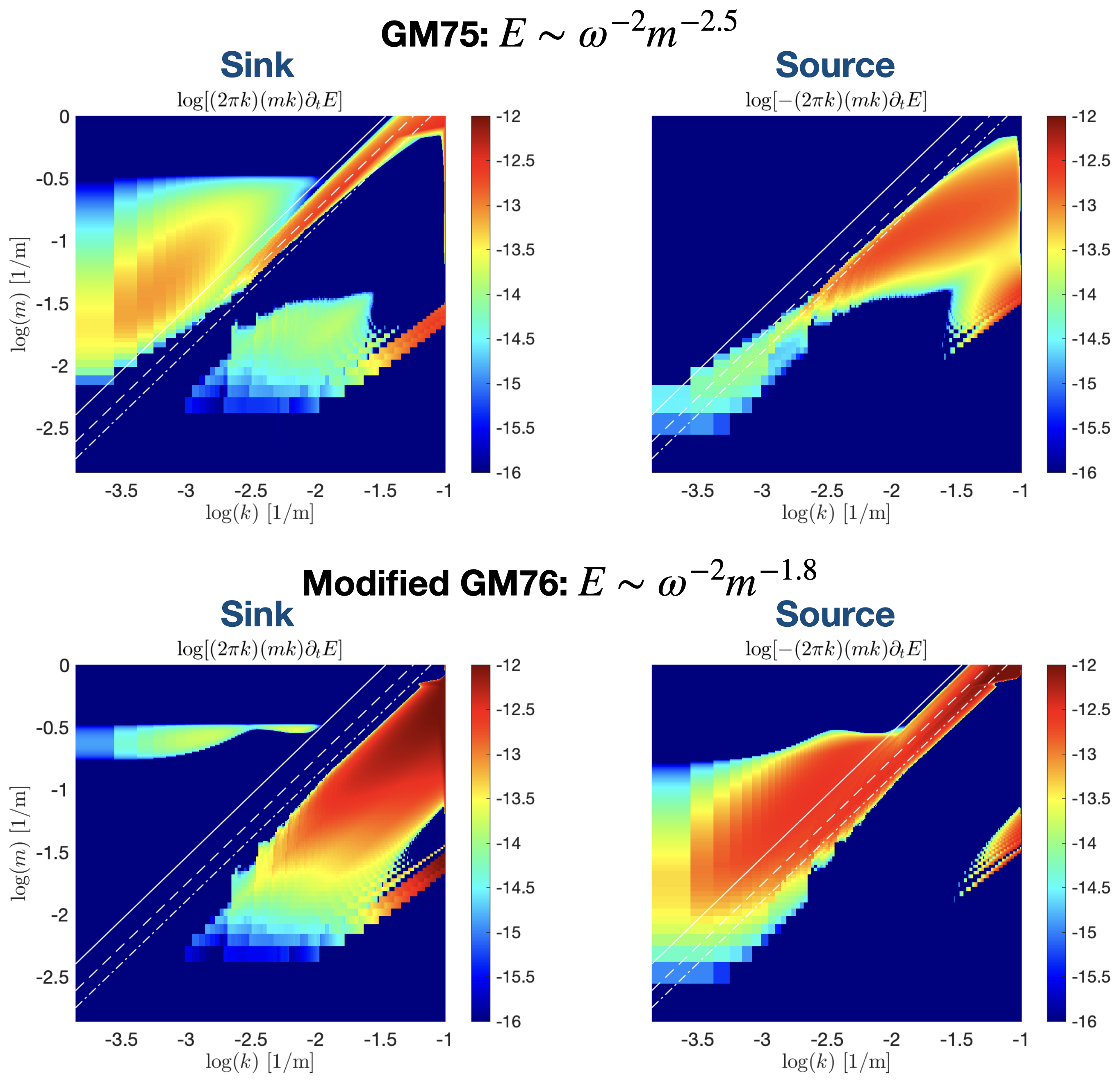}}
 \caption{As Figure~\ref{fig_Et} but for the induced diffusion mechanism. (Upper) GM75 spectrum with $E(\omega,m)\sim \omega^{-2} m^{-2.5}$, and (lower) a modified GM76 spectrum with $E(\omega,m)\sim \omega^{-2} m^{-1.8}$.}
 \label{fig_Et_slope}
\end{figure}

We can further deduce the ID dynamics under conditions $q>0$ and $q<0$. For $q>0$ and $q<0$ respectively, the action diffuses toward higher and lower $m$ (with the same $k$), indicating a backward and a forward cascade in frequency. We can leverage our numerical tools to verify these inferences. For the former with $q>0$, we consider a GM75 spectrum with $p=0$ and $q=0.5$. The energy transfer due to ID is computed as shown in Figure~\ref{fig_Et_slope}a, where we do see a dominating backward cascade in frequency. For the latter, we consider a realistic spectrum $E(\omega,m)\sim \omega^{-2}m^{-1.8}$ with $p=0$ and $q=-0.2$ taken from field measurements in the Southeast Subtropical North Pacific \citep{Polzin2011} reporting $E(\omega,m)\sim \omega^{-2}m^{-1.9 \sim -1.75}$. The ID energy transfer, as shown in Figure~\ref{fig_Et_slope}b, indeed exhibits a dominating forward cascade.
In summary, the ID mechanism can lead to different directions of energy cascade depending on the spectral slopes of the IGW continuum and should be understood with respect of the specific spectrum of interest.

\subsubsection*{\textbf{The local interactions}}
The energy transfer by local interactions is shown in Figure~\ref{fig_Et_sep}d. We see a clear bi-directional cascade in frequency extracting energy out of the frequency band $[2f,4f]$. The transfer is not only non-negligible but stronger than any other interaction mechanisms. McComas~\textit{et~al.}'s early assumption about the dominance of scale-separated triad interactions in energy transfer is clearly incorrect for the GM76 (and perhaps general) spectrum. While it is difficult to exactly trace the ground based on which McComas~\textit{et~al.} made this assumption, it is likely related to some simple calculation regarding the interaction coefficient $V$ and red IGW action spectra. However, a comprehensive understanding of the problem also relies on other factors, such as the number of triads participating in energy transfer and the specific form of the GM76 action spectrum.
Our direct calculation of \eqref{eq_KE} encapsulating all factors clearly demonstrates the paramount importance of local interactions for the GM76 spectrum. The result here also echos those in \cite{Dematteis2021} and \cite{Dematteis2022} for a modified GM76 spectrum in the scale-invariant case.

\subsection{Constituents of energy flux}
The energy flux in directions $k$, $m$ and $\omega$ due to each mechanism are plotted in Figure~\ref{fig_P}. For forward cascade in $k$, we see that majority of the cascade is provided by local interactions, with ES contributing a relatively small fraction. For forward cascade in $m$, local interactions and PSI each contribute approximately half to the total flux. The energy transfer in frequency exhibits a bi-directional flux. The backward flux, moving energy from $[2f, 4f]$ to lower frequencies, is supplied by both PSI and local interactions with similar magnitudes. The forward cascade results from local interactions, ES and ID (mainly from sub-diffusion process) with descending contributions. Among all directional cascades, local interactions are the only mechanism participating significantly in all of them, which was neglected in McComas~\textit{et~al.}'s early works.

We are eventually in a good position to state our new understanding regarding paradoxes (a) and (b). For (a), we now understand that the frequency cascade is bi-directional, with the forward flux provided by local interactions, ES and ID, all elements ignored in McComas~\textit{et~al.}'s works. For (b), the ID mechanism in GM76 is indeed approximately a zero-flux state, except forming a weak forward cascade in frequency through the sub-diffusion process. McComas~\textit{et~al.}'s argument about ID providing significant portion of downscale flux should be replaced by local interactions.

\section{Conclusions and discussions}
Through direct evaluation of the collision integral in WKE of IGWs, we study the spectral energy transfer for the GM76 spectrum. Our calculation of the downscale energy flux, through its maximum value over all vertical wavenumbers, provides an estimate in close agreement with the finescale parameterization. We also analyse different interaction mechanisms, resolving some long-standing paradoxes in the field. Our new understanding includes: 
\begin{enumerate}[leftmargin=.6cm,labelsep=.1cm,label={(\arabic*)}]
\item \noindent Local interactions are important for energy cascade in all spectral directions, which were completely neglected in early works by McComas~\textit{et al.}.
\item \noindent The downscale energy flux (toward high vertical wavenumbers) is supplied by PSI and local interactions, rather than PSI and ID as understood in McComas~\textit{et~al.}'s works. 
\item \noindent ID can provide cascade toward different directions depending on spectral slopes of the IGW continuum. For GM76, the leading-order flux by ID vanishes, with sub-diffusion process providing a weak forward frequency cascade. 
\item \noindent The ES mechanism provides a forward frequency cascade (but no cascade in wavenumbers) in vertically symmetric IGW fields, which was not investigated in previous works.
\end{enumerate}

Our capability of numerical evaluation of WKE opens a new door to an advanced understanding of oceanic IGW cascade and mixing. Among all possible directions of future study, an immediately fruitful one is to evaluate the flux properties and magnitudes for various IGW spectral forms. As revealed in field measurements (e.g., \citealt{Polzin2011}) and wave turbulence theory (e.g., \citealt{Lvov2010} in terms of stationary solutions to the WKE), the oceanic IGW spectrum exhibits large variability, deviating from the GM76 model. Under this circumstance, it is not clear whether the current finescale formula \eqref{finescale}, developed mainly referencing to the GM76 spectrum, is sufficiently robust for all IGW spectral forms. Our numerical method provides direct approach through which this problem can be studied. In addition, combining our WKE predictions with recent high-resolution regional ocean simulations \cite[e.g.][]{Nelson2020,Thakur2022,Skitka2023} may bring new insights to the field.

We also would like to mention that there is an active debate on the relative importance between the nonlinear wave-wave interactions and wave-eddy interactions in IGW energy cascade, and our current paper clearly does not consider the latter. In order to consider both processes, it may be beneficial, as a first-order approximation, to include an additional eddy-diffusion term \cite[][]{Kafiabad2019,Dong2020} in the WKE and study the full equation (note that this eddy diffusion is linear assuming stationary eddy field). While these are future directions we plan to consider, the current work perhaps already has shed some light on the problem in terms of understanding the importance of local interactions that is only part of the wave-wave interactions.

\subsubsection*{\textbf{Funding}}
This research is supported by the National Science Foundation (award number OCE-2241495 and OCE-2306124) and the Simons Foundation through Simons Collaboration on Wave Turbulence.

\subsubsection*{\textbf{Acknowledgement}}
The authors thank Giovanni Demantteis, Yuri Lvov, Sergey Nazarenko, Jalal Shatah, Brian Arbic, Joseph Skitka and two anonymous reviewers for their helpful advice. 

\subsubsection*{\textbf{Data availability statement}}
The data that support the findings of this study are openly available on GitHub at https://github.com/yue-cynthia-wu.

\subsubsection*{\textbf{Declaration of interests}}
The authors report no conflict of interest.


\appendix


\section{Reduction of kinetic equation on resonant manifold}

For a horizontally isotropic IGW spectrum, it is convenient to convert the collision integral to cylindrical coordinates and integrate out the dependence on horizontal azimuth. For example, the term of summation interactions can be transformed to 
\begin{eqnarray}  \label{eq:A+1}
I^+
&&=\int\!\!\!\!\!\iiiint\displaylimits_{-\infty}^{+\infty}\!\!\!\!\!\int J^+\,dk_{1x}\,dk_{1y}\,dk_{2x}\,dk_{2y}\,dm_1\,dm_2 \\
&&= \int\displaylimits_{-\infty}^{+\infty} \int\displaylimits_{-\infty}^{+\infty} \int\displaylimits_0^{2\pi} \int\displaylimits_0^{2\pi} \int\displaylimits_0^{+\infty} \int\displaylimits_0^{+\infty}  k_1 k_2 J^+\,dk_1\,dk_2\,d\theta_1\,d\theta_2\,dm_1\,dm_2,
\end{eqnarray}
where $J^+=4\pi |V^{\mathbf{p}}_{\mathbf{p}_1,\mathbf{p}_2}|^2 \, f_{p12} \, \delta(\omega-\omega_1-\omega_2) \delta(\mathbf{k}-\mathbf{k_1}-\mathbf{k_2}) \delta(m-m_1-m_2)$, 
$\mathbf{p}=(\mathbf{k},m)=(k_x,k_y,m)$ is the three-dimensional wavenumber vector, $\mathbf{k}=(k_x,k_y)$ is the horizontal wavenumber vector, and $k=|\mathbf{k}|$ is the horizontal wavenumber magnitude.

We define
\begin{eqnarray}
Q^+= 4\pi |V^{\mathbf{p}}_{\mathbf{p}_1,\mathbf{p}_2}|^2 \, f_{p12} \, \delta(\omega-\omega_1-\omega_2)\delta(m-m_1-m_2) k_1 k_2.
\end{eqnarray}

Thus
{\scriptsize
\begin{eqnarray} \label{eq:A+}
I^+&&=\int\displaylimits_{-\infty}^{+\infty} \int\displaylimits_{-\infty}^{+\infty} \int\displaylimits_0^{+\infty} \int\displaylimits_0^{+\infty} \int\displaylimits_0^{2\pi} \int\displaylimits_0^{2\pi} Q^+\delta(\mathbf{k}-\mathbf{k_1}-\mathbf{k_2})\,d\theta_1\,d\theta_2\,dk_1\,dk_2\,dm_1\,dm_2\\ \nonumber
&&=\int\displaylimits_{-\infty}^{+\infty} \int\displaylimits_{-\infty}^{+\infty} \int\displaylimits_0^{+\infty} \int\displaylimits_0^{+\infty}\int\displaylimits_0^{2\pi} \int\displaylimits_0^{2\pi} Q^+\delta(k-k_1\cos \theta_1 -k_2\cos \theta_2)\delta(k_1\sin \theta_1+k_2\sin \theta_2)\,d\theta_1\,d\theta_2\,dk_1\,dk_2\,dm_1\,dm_2 \\\nonumber
&&=\int\displaylimits_{-\infty}^{+\infty} \int\displaylimits_{-\infty}^{+\infty} \int\displaylimits_0^{+\infty} \int\displaylimits_0^{+\infty} 2Q^+_{\triangle^+}/|\mathbcal{J}|\,dk_1\,dk_2\,dm_1\,dm_2 \\\nonumber
&&=\int\displaylimits_{-\infty}^{+\infty} \int\displaylimits_{-\infty}^{+\infty} \int\displaylimits_0^{+\infty} \int\displaylimits_0^{+\infty}Q^+_{\triangle^+}/S_{\triangle}\,dk_1\,dk_2\,dm_1\,dm_2,
\end{eqnarray}
}
where $\mathbcal{J}$ is the Jacobian, and $S_{\triangle}$ is the area of the triangle formed by $\mathbf{k}$, $\mathbf{k_1}$ and $\mathbf{k_2}$. 
The subscript $\triangle^+$ of $Q^+$ denotes the projection on the manifold of $\mathbf{k}=\mathbf{k_1}+\mathbf{k_2}$ for given $k$, $k_1$ and $k_2$. 
The angle integration involved in the above equation can be found in many wave turbulence literatures, e.g., \cite{Zakharov1992, Lvov2012, Pan2017b}.

Repeating this procedure for reduction interactions, we obtain
\begin{eqnarray}
Q^-= 4\pi |V^{\mathbf{p_1}}_{\mathbf{p},\mathbf{p}_2}|^2 \, f_{1p2} \, \delta(\omega-\omega_1+\omega_2)\delta(m-m_1+m_2) k_1 k_2,
\end{eqnarray}

\begin{eqnarray} \label{eq:A-}
I^-=\int\displaylimits_{-\infty}^{+\infty} \int\displaylimits_{-\infty}^{+\infty} \int\displaylimits_0^{+\infty} \int\displaylimits_0^{+\infty} Q^-_{\triangle^-}/S_{\triangle}\,dk_1\,dk_2\,dm_1\,dm_2,
\end{eqnarray}
where the subscript $\triangle^-$ of $Q^-$ denotes the projection on the manifold of $\mathbf{k_1}=\mathbf{k}+\mathbf{k_2}$ for given $k$, $k_1$ and $k_2$. 

\vspace{1cm}
We further integrate out the delta function in vertical wavenumbers in \eqref{eq:A+} and \eqref{eq:A-}, which gives
{\footnotesize
\begin{eqnarray}
CL= I^+ + 2I^- =\iint_0\displaylimits^{+\infty}
   \left\{\int\displaylimits_{-\infty}^{+\infty} h^+(k, k_1, k_2, m, m_1) \delta\left[g^+(k, k_1, k_2, m, m_1)\right] dm_1 \right.\\ \nonumber
-2 \left. \int\displaylimits_{-\infty}^{+\infty} h^-(k, k_1, k_2, m, m_1) \delta\left[g^-(k, k_1, k_2, m, m_1)\right] dm_1 \right\}\,dk_1\,dk_2.
\end{eqnarray}
}

In the inside integral, for given $k$, $m$, $k_1$ and $k_2$,
\[h^+(m_1) = 4\pi |V^{\mathbf{p}}_{\mathbf{p}_1,\mathbf{p}_2}|^2_{\triangle^+,m^+} \, f_{p12,\triangle^+,m^+} \, k_1 k_2/S_{\triangle},\]
\[h^-(m_1) = 4\pi |V^{\mathbf{p}_1}_{\mathbf{p},\mathbf{p}_2}|^2_{\triangle^-,m^-} \, f_{1p2,\triangle^-,m^-} \, k_1 k_2/S_{\triangle},\]
\[g^+(m_1) = \omega-\omega_1(m_1)-\omega_2(m-m_1),\]
\[g^-(m_1) = \omega-\omega_1(m_1)+\omega_2(m_1-m),\]
where the subscript $m^+$ ($m^-$) of the interaction coefficient $V$ and quadratic function $f_{p12}$ ($f_{1p2}$) denotes projection on $m=m_1+m_2$ ($m_1=m+m_2$).

Finally, $CL$ is reduced to an integration over two dimensions $k_1$ and $k_2$ by integrating out delta function in frequency
\begin{equation}
CL= \iint_0\displaylimits^{+\infty} \left[\frac{h^+(m^{*+}_1)}{|{g^+}'(m^{*+}_1)|}-2\frac{h^-(m^{*-}_1)}{|{g^-}'(m^{*-}_1)|}\right] \,dk_1\,dk_2,
\label{CL_F}
\end{equation}
where $m^{*+}_1$ is the root of $g^+(m_1)=0$ for summation interactions, and $m^{*-}_1$ is the root of $g^-(m_1)=0$ for reduction interactions. 
The denominators ${g^+}'(m_1)$ and ${g^-}'(m_1)$ are the $m_1$-derivatives of functions $g^+(m_1)$ and $g^-(m_1)$, respectively [also see \citealt{Eden2019b}, eq. (8)]
\begin{equation}
{g^+}'(m_1)={g^-}'(m_1)=\frac{m_1}{\omega_1}\frac{\omega_1^2-f^2}{k_1^2+m_1^2}-\frac{m_2}{\omega_2}\frac{\omega_2^2-f^2}{k_2^2+m_2^2}.
\end{equation}

\vspace{1cm}
We note that \eqref{eq:A+} and \eqref{eq:A-} involve a singularity (first-order pole) when $\mathbf{p}$, $\mathbf{p_1}$ and $\mathbf{p_2}$ lie on the same vertical plane, i.e., $k=k_1+k_2$ or $k_1=k+k_2$, leading to $S_{\triangle}=0$ and corresponding to the collinear triad interactions described in \cite{Dematteis2021}. However, this is an integrable singularity since it is not involved in \eqref{eq:A+1} before the coordinate transform and angle integration. Therefore, given sufficiently fine grid resolution of $k_1$ and $k_2$, the singularity point can be neglected in the integration of \eqref{CL_F}. For finite grid resolution, we can approximately retrieve the contribution from the nearby region of the singularity point using \eqref{eq:A+1} (and its counterpart for reduction interactions). Specifically, for given $k$ and $m$, contributions from vicinity of all singularities in the integral of \eqref{CL_F} can be accounted for by treating \eqref{eq:A+1} with the following procedures: set (due to isotropy) $\mathbf{k}=(k,0)$; integrate out the delta function in $\mathbf{k}$ with respect to the integration over $k_{2x}$ and $k_{2y}$; consider $\mathbf{k}_1$ in the same direction as $\mathbf{k}$ by setting $k_{1y}=0$ and convert the integration along $k_{1y}$ into the integral multiplying by the grid size $\Delta k$; change the dummy variable $k_{1x}$ into $k_1$; integrate out the $\omega$ and $m$ delta functions with respect to the integration over $m_1$ and $m_2$. These procedures lead to
\begin{equation}
CL_0= \Delta k \int_0\displaylimits^{+\infty} \left[\frac{h_0^+(m^{*+}_1)}{|{g^+}'(m^{*+}_1)|}-2\frac{h_0^-(m^{*-}_1)}{|{g^-}'(m^{*-}_1)|}\right] \,dk_1,
\end{equation}
\[h_0^+(m_1) = 4\pi |V^{\mathbf{p}}_{\mathbf{p}_1,\mathbf{p}_2}|^2_{\parallel^+,m^+} \, f_{p12,\parallel^+,m^+},\]
\[h_0^-(m_1) = 4\pi |V^{\mathbf{p}_1}_{\mathbf{p},\mathbf{p}_2}|^2_{\parallel^-,m^-} \, f_{1p2,\parallel^-,m^-}.\]
where the subscript $\parallel^+$ denotes projection on $\mathbf{k}=(k,0)$, $\mathbf{k}_1=(k_1,0)$ and $\mathbf{k}_2=(k-k_1,0)$, and $\parallel^-$ projection regarding the reduction interactions. 


\section{Root finding for $m^{*+}_1$ and $m^{*-}_1$}
\begin{figure} \label{fig_root-finding}
  \centerline{\includegraphics[width=\textwidth]{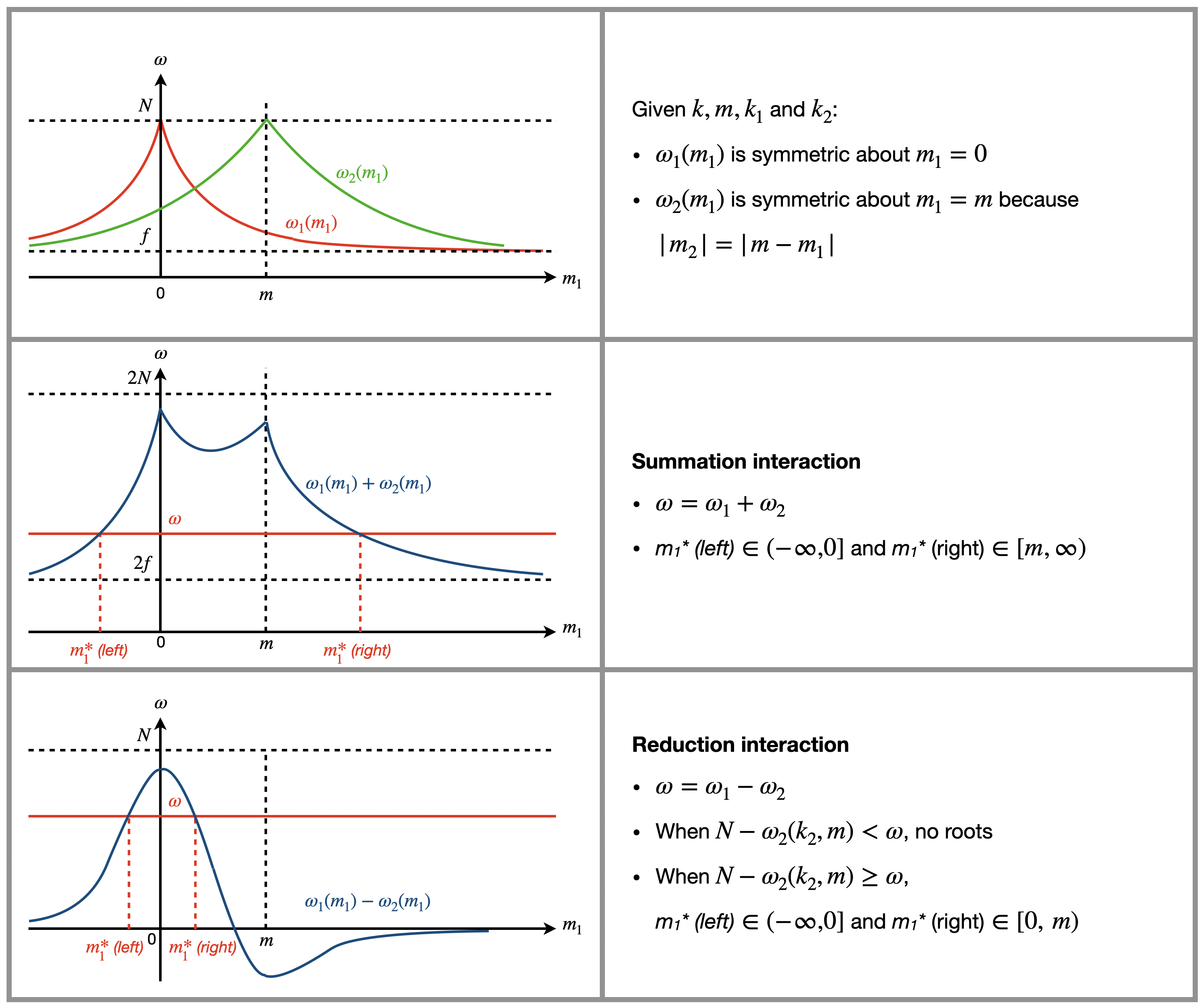}}
  \caption{Graphical interpretation on the root finding of $m^*_{1+}$ and $m^*_{1-}$ for an example case given $k$, $k_1$, $k_2$ and $m$.}
\end{figure}

For given $k$, $k_1$, $k_2$ and $m$, we search for the roots $m^{*+}_1$ and $m^{*-}_1$ of $g^+(m_1)=g^-(m_1)=0$ for summation and reduction interactions, respectively.
Figure~\ref{fig_root-finding} shows a graphical interpretation on the root finding for an example case. In this example, it is shown that two roots of $m^{*+}_1$ exist for summation interactions and two roots of $m^{*-}_1$ exist for reduction interactions. We note that this example of geometric interpretation cannot rule out the situation where different number of roots exist for either type of interactions, especially for finite range of $m$ as in the simulation. In order to cover all possible roots in our numerical root-finding procedure, we set two initial guesses $m_1=(0,m)$ and start searching for the roots toward the left and right of both initial guesses along the $m_1$-axis using Brent's algorithm \citep{brent1971} until covering the full range of the numerical spectral domain.


\bibliographystyle{jfm}
\bibliography{igw}

\end{document}